\documentclass[%
 preprint, linenumbers,
 amsmath,amssymb,
 aps, physrev,
]{revtex4}

\usepackage{graphicx}
\usepackage{dcolumn}
\usepackage{bm}

\begin{document}

\title{Halo-dependent Anharmonic Effects in Collective Excitation for Light Dark Matter Direct Detection}

\author{Jun Guo}
\thanks{\href{mailto:jguo\_dm@jxnu.edu.cn}{jguo_dm@jxnu.edu.cn}}
\affiliation{College of Physics and Communication Electronics,
Jiangxi Normal University, Nanchang 330022, China}

\author{Lei Wu}
\thanks{\href{leiwu@njnu.edu.cn}{leiwu@njnu.edu.cn}}
\affiliation{Department of Physics and Institute of Theoretical Physics, Nanjing Normal University, Nanjing, 210023, China}

\author{Bin Zhu}
\thanks{\href{mailto:zhubin@mail.nankai.edu.cn}{zhubin@mail.nankai.edu.cn}}
\affiliation{Department of Physics, Yantai University, Yantai 264005, China}

\begin{abstract}
Phonon, the collective excitation of lattice vibration in the crystal, has been put forward as a means to search for light dark matter. However, the accurate modeling of the multi-phonon production process is challenging in theory. The anharmonicity of the crystal must be taken into account, as it has a significant impact on dark matter-nucleus scattering cross section in the low dark matter mass region. Notably, such an effect is sensitive to the velocity distribution of the dark matter halo. In this work, we consider the potential dark matter substructures indicated by the recent Gaia satellite observation and investigate their impact on the anharmonicity of the silicon crystal. By employing the likelihood analysis with the Asimov dataset, we present the expected sensitivity of dark matter-nucleus interactions, which can differ from the standard halo model by a factor of 2-3.

\end{abstract}
\maketitle

\section{Introduction}
\label{sec:intro}
Various observations from cosmology and astrophysics indicate that approximately $80\%$ of the matter mass is constituted of dark matter (DM). It is now believed that the Milky Way is embedded and rotates in a spherical DM halo. However, the microscopic nature of DM remains a mystery in particle physics. Over the past few years, the sensitivity of direct detection experiments aimed at Weakly Interacting Massive Particles (WIMPs) has reached the neutrino floor, yet no substantial signals have been identified. This necessitates the investigation of alternative scenarios. Consequently, sub-GeV dark matter has emerged as an increasingly intriguing candidate for DM studies. For the light DM with mass $m_{\chi}\sim\mathcal{O}(100)$ MeV, the deposited energy of the DM-nucleus scattering process is $E_r \sim q^2/m_N \sim 0.1$ keV, which is much lower than the thresholds of the typical noble liquid detectors. On the other hand, there have been great efforts devoted to the targets with eV excitation energies, such as silicon and germanium. The delicate effects of atoms~\cite{Essig:2011nj,Graham:2012su,Lee:2015qva,Essig:2015cda,Essig:2017kqs,Catena:2019gfa}, ions and electrons in the condensed matter systems\cite{Derenzo:2016fse,SuperCDMS:2018mne,Kurinsky:2019pgb,SENSEI:2019ibb,DAMIC:2019dcn,Trickle:2019nya,Griffin:2019mvc,Andersson:2020uwc,SENSEI:2020dpa,Catena:2021qsr,Flambaum:2020xxo, Wang:2021nbf,Wang:2021oha,Gu:2022vgb,Li:2022acp,Su:2022wpj,Gu:2023pfg,Liang:2024xcx}  were studied. 

As the collective excitation of lattice vibration in the crystal, phonon has been put forward as a means to search for light dark matter. Lots of previous works have provided a calculation process with single phonon excitation, but it is more plausible that the multi-phonon final state will occur. In the~\cite{Kahn:2020fef}, the authors provide the analytic calculation process for the multi-phonon final state in the one dimensional approximation, with the phonon occupation number following a Poisson distribution. When the nucleus recoil energy $E_r$ is smaller than the replacement energy $E_d\sim \mathcal{O}(10)\, \rm{keV}$ in the lattice, the nucleus can be approximated as a harmonic oscillator. Nevertheless, the anharmonicity of the crystal cannot be neglected as it significantly affects the scattering rate~\cite{Lin:2023slv}. The impact is substantial, potentially leading to a variation in the scattering rate by two orders of magnitude. Particularly, the anharmonic effect will become important at the low transfer momentum and high transfer energy region. This is directly correlated with the DM velocity distribution around the Earth, which is also a key factor in direct detection.

Additionally, the recent observation from the Gaia satellite ~\cite{Gaia:2016lbo,Gaia:2018ydn,Gaia:2016zol} have revealed that the stellar halo in the vicinity of the solar system is imprinted by a rich variety of substructures. It indicates that each stellar substructure has a corresponding DM substructure. These will make a great difference in the DM velocity distributions and lead to a considerable impact on the anharmonic effect and direct detection. In this paper, we investigate the DM-nucleus scattering in the crystal by including the anharmonic effect for different DM velocity distributions. The paper is organized as follows: in Section 2, we recapitulate the calculations of the phonon signature in DM-nucleus scattering for a crystal target and discuss the anharmonic effects and DM substructures and their impact on the scattering rate. Section 3 details the data analysis methodology and presents the expected sensitivity in the direct detection. Finally, we summarize our conclusions in Section 4.

\section{Dark Matter-Nucleus Scattering in the Crystal}

\subsection{Anharmonic Effect}

When the ambient DM scatters off the nucleus in the crystal, the process differs significantly from its scattering off the free nucleus, such as Xenon. In the crystal, nuclei are bounded within a potential formed by neighboring atoms. When the transferred energy of the scattering is smaller than the binding energy of the lattice, the available degrees of freedom of the nucleus are phonons. To understand this process, the most critical term is the structure factor ~\cite{Campbell-Deem:2022fqm,Lin:2023slv}, which represents the linear response of the crystal under dark-matter scattering. For a given transferred energy $\omega$ and momentum $\textbf{q}$, it is
\begin{align}
        S(\textbf{q},\omega) = \frac{2\pi}{V}\sum_f  \Big|\sum_{\boldsymbol{\ell}}\sum_{d}f_{\boldsymbol{\ell} d}\langle \Phi_f|e^{i\textbf{q} \cdot \textbf{r}_{\boldsymbol{\ell}d}}|0\rangle\Big|^2 \delta(E_f - E_0 -\omega).
\end{align}
where $V$ is the total volume of the system, $f_{\boldsymbol{\ell} d}$ is the lattice coupling, with $\boldsymbol{\ell}$ being the lattice vectors of the cell and $d$ indexing the atoms in the cell, $\textbf{r}_{\boldsymbol{\ell}d}$ is the position of the target nucleus, and $|\Phi_f\rangle$ and $E_f$ are the excited final state and energy, corresponding to the phonon eigenstates and eigenenergies of the lattice system.

To analytically calculate $S(\textbf{q},\omega)$, two approximations are applied. The first is the incoherent approximation, which assumes that the transfer momentum $q\leq 1/a$, where $a$ is the inter-atomic distance. Consequently, the scattering is localized at a single lattice site, and the structure factor simplifies to:
\begin{align}
        S(\textbf{q},\omega) = \frac{2\pi}{V}\sum_f \sum_{\boldsymbol{\ell}}\sum_{d} |f_{\boldsymbol{\ell} d}|^2\Big|\langle \Phi_f|e^{i\textbf{q} \cdot \textbf{r}_{\boldsymbol{\ell}d}}|0\rangle\Big|^2  \delta(E_f - E_0 -\omega).
\end{align}
without interference terms between atoms. 

Another approximation assumes that the final states are isolated atomic states, meaning that each atom is in an isolated potential. By replacing the delta function of energy conservation with a Gaussian distribution, the structure factor is expressed as,

\begin{align}
    S(q,\omega) = 2\pi \sum_{d} n_d |f_d|^2 \sum_f \Big|\langle \Phi_f|e^{iqx }|0\rangle\Big|^2  \frac{1}{\sqrt{2\pi f(n) \sigma^2}} e^{- \frac{(\omega - f(n) \omega_0)^2}{2 f(n) \sigma^2}} \times \Theta(\omega_{\mathrm{max}} - \omega).
    \label{eq:sfn}
\end{align}
in which
\begin{align}
    \omega_0 &= \Big( \int d \omega \omega^{-1} D(\omega) \Big)^{-1},
    \\
    \sigma &= \sqrt{\frac{\int d\omega \omega D(\omega)}{\omega_0} - \frac{1}{\omega_0^2}},    \\
    \omega_\mathrm{max} &= f(n) \times (\mathrm{min}(\omega) \vert D(\omega) = 0 )
\end{align}

In the above expressions, $D(\omega)$ represents the state density of a single phonon~\cite{Knapen:2021bwg}. The term $f(n)\omega_0$ denotes the energy difference between the nth eigenenergy and the ground state. 

To obtain the matrix element $\langle \Phi_f|e^{iqx }|0\rangle$ analytically, we treat the potential as a harmonic one, following ~\cite{Kahn:2020fef,Lin:2023slv}, the matrix element square is,

\begin{align}
    \Big|\langle \Phi_f|e^{iqx }|0\rangle\Big|^2 = \frac{1}{n!} \left(\frac{q^2}{2 m_d \omega_0} \right)^n \exp\left(-\frac{q^2}{2 m_d \omega_0}\right)
\end{align}
which is a Poisson distribution with rate $\lambda={q^2}/{2 m_d \omega_0}$. However, calculating the structure factor with greater precision requires accounting for the anharmonicities of the crystal potential. This means that the potential is not solely characterized by a quadratic term; higher-order terms, such as cubic and quartic terms, must also be included. 

Concurrently, the Morse potential is a natural description of anharmonicity in the molecular vibrations. Furthermore, it is a physical model that permits exact solutions. The form of Morse potential is defined as follows,

\begin{equation}
    V_{\mathrm{Morse}} = \frac{\omega_0}{64 \lambda_M^2} \Big( e^{8 \lambda_M \sqrt{2 m_d \omega_0} x} - 2 e^{4 \lambda_M \sqrt{2 m_d \omega_0} x} \Big).
\end{equation}
The anharmonicity is described by the parameter $\lambda_M$, with the width of the potential controlled by the value $4\lambda_M\sqrt{2m_d\omega_0}$. Given that the Morse potential has only one parameter $\lambda_M$, it is only capable of describing the cubic term. However, this precision is sufficient for the study of anharmonicity in crystals. 

The matrix element square under the Morse potential is~\cite{berrondo1987}:
\begin{align}
    \Big|\langle \Phi_f|e^{iqx }|0\rangle\Big|^2 = \frac{(2K-2n-1)(2K-1)}{n!\Gamma(2K) \Gamma(2K-n)}  \Big|\frac{\Gamma (n + \frac{i(q/\sqrt{2m_d \omega_0})}{4\lambda_M}) \Gamma(2K + \frac{i(q/\sqrt{2m_d \omega_0})}{4\lambda_M} - n - 1)}{\Gamma(\frac{iq/\sqrt{2m_d \omega_0}}{4\lambda_M})} \Big|^2.
    \label{eq:morse_eiqx}
\end{align}
The energy gap between the n-th eigenenergy and the ground state energy is
\begin{align}
E_n - E_0 = \left(n - \frac{n(1+n)}{2K}\right)\omega_0
\end{align}
with $K=\frac{1}{32\lambda_M^2}$ and $\lambda_M=0.02$. To keep the final state as the bounded state, it is required that $n<K-1/2$. The material we consider is the silicon crystal, with $\omega_0=30.8\,\rm{meV}$, $\sigma=17.6\,\rm{meV}$ and $\sqrt{2m_d\omega_0}=40.3\,\rm{keV}$ 

\begin{figure*}[htbp]
\centering
\includegraphics[width=0.45\textwidth]{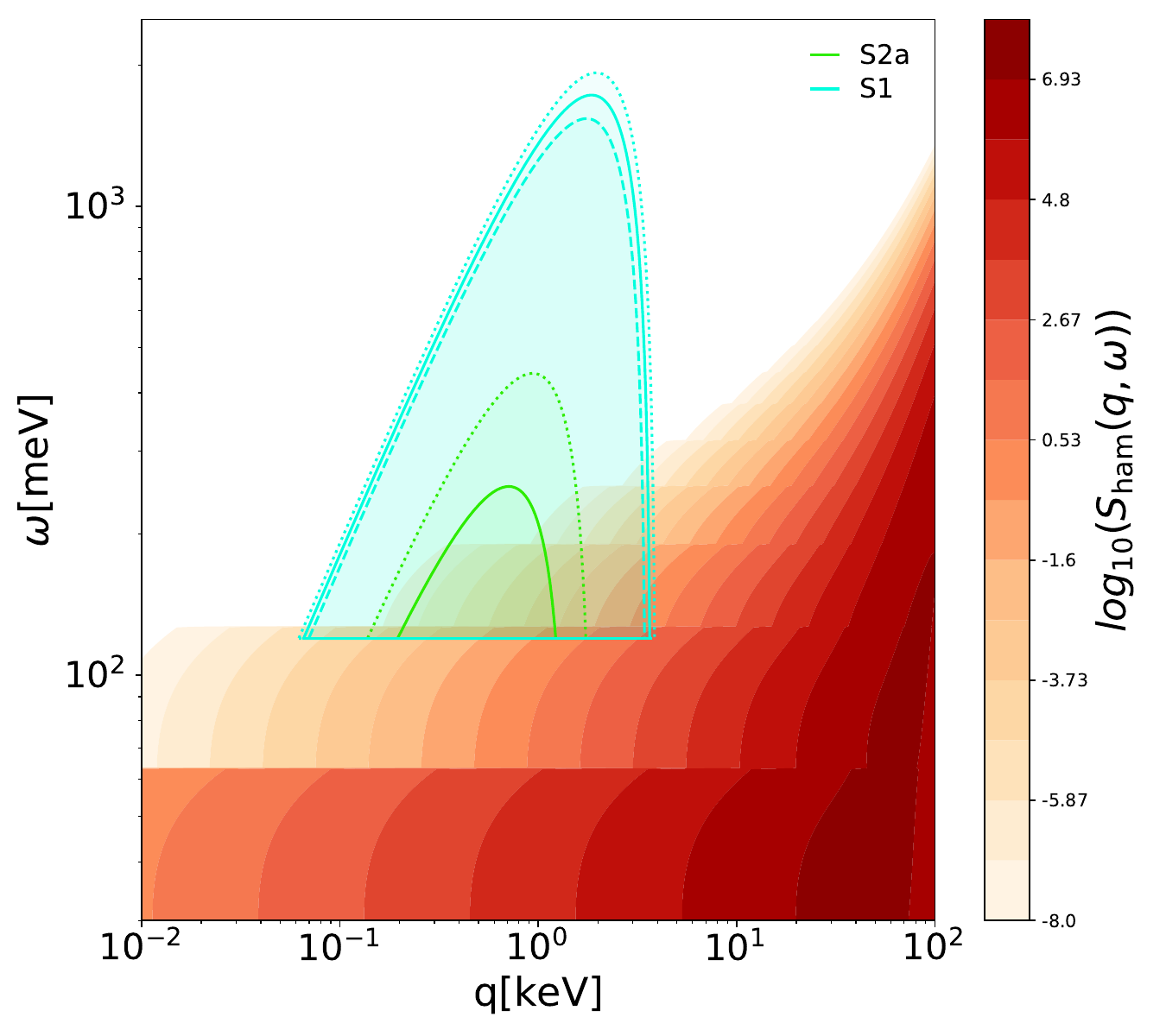}
\includegraphics[width=0.45\textwidth]{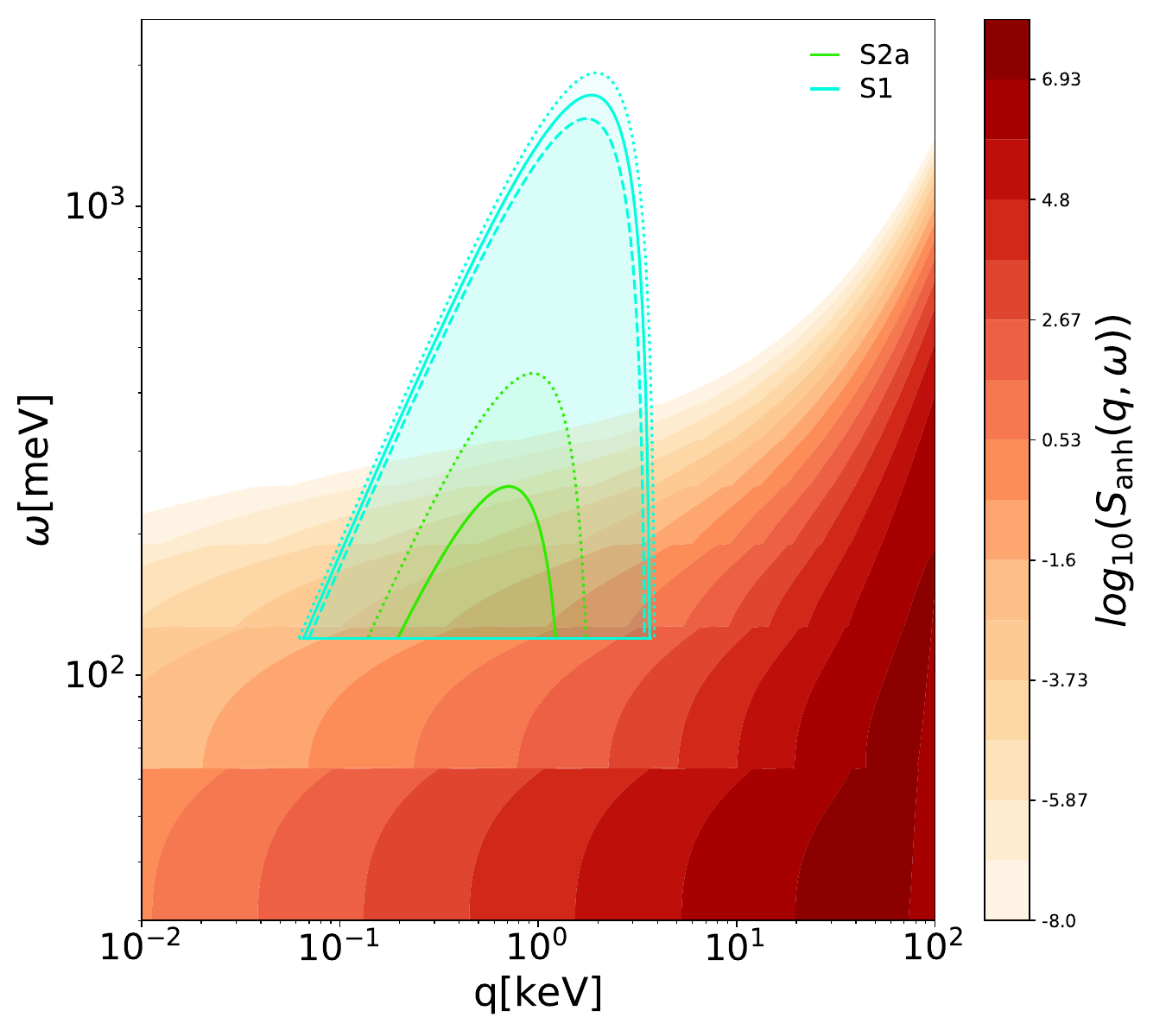}
\caption{Left: the structure factor function for the harmonic case. Right: the structure factor function for the anharmonic case. The anharmonic structure factor function prefers to have a non-negligible value at the low transfer momentum high transfer energy region. We also show the momentum-energy integral range for the S2a and S1 stream, with the solid line representing the range for $v_{mp}$, and the dotted 
 dashed lines represent the upper $v_{up}$ and lower $v_{low}$ bound of 1-$\sigma$ region around $v_{mp}$.}
 \label{fig:S_plot}
\end{figure*}

In the Fig.~\ref{fig:S_plot}, we show the plot for structure factor function $S(q, \omega)$ for the harmonic and anharmonic cases. As $\omega$ becomes larger, the $S(q, \omega)$ value is exponentially suppressed from Eq.~\ref{eq:sfn}. For a larger $\omega$, the difference of structure factor function value between harmonic and anharmonic cases becomes more significant. When the transfer momentum $q\ll q_0$ is small, the anharmonic effect is significant, meaning an increase in the allowed phonon number. When the transfer momentum $q$ becomes larger, the difference of $S(q, \omega)$ between two cases disappears.

\subsection{Velocity Distribution Impact on Anharmonic Effect}

In light of the structure factor outlined in the above section, the event rate of phonon scattering can be calculated as follows,
\begin{align}
    R =  \frac{1}{4\pi \rho_T} \frac{\rho_\chi}{m_\chi}  \frac{\sigma_p}{\mu_\chi^2}  \int \! d^3 \textbf{v}\, \frac{f_{\rm{lab}}(\textbf{v})}{v} \int\displaylimits_{q_-}^{q_+} \! dq \,  \int\displaylimits_{\omega_\mathrm{th}}^{\omega_+} \! d\omega \,   q \, F^2_{\rm{DM}}(q) S(q, \omega),
    \label{Eq:rate}
\end{align}
where $\rho_T$ is the target material mass density, $\rho_\chi$ is the local dark matter density $0.4\,\rm{GeV}/\rm{cm}^3$, $\sigma_{\rm p}$ and $\mu_\chi$ are the DM-nucleus scattering cross section and reduced mass. The DM form factor $F(q)$ encodes all particle-physics-related information. It is typically set to $1$ for the "heavy mediator" case or $q_0^2/q^2$ for the light mediator case, where $q$ is the momentum transfer and $q_0=\sqrt{2m_d\omega_0}$. The lower bound of the energy integration is the experimental threshold $\omega_{\rm{th}}$, and the upper bound $\omega_{up}$ corresponds to the maximum allowed energy transfer,
\begin{align}
    \omega_{up}=qv - q^2/2m_\chi
    \label{eq:int_omg}
\end{align}
where $v$ is the DM speed. The upper and lower limits of the $q$-integration $q_{\pm}$ are given by
\begin{align}
  q_{\pm}=m_\chi v(1\pm\sqrt{1-\omega_{\rm{th}}/m_\chi v^2}) 
  \label{eq:int_q}
\end{align}

The function $f_{\rm{lab}}(\textbf{v})$ represents the distribution of dark matter (DM) velocities in the laboratory frame. It contains comprehensive information about the DM halo and significantly influences the scattering event rate. The velocity distribution $f_{\rm{lab}}(\textbf{v})$ is derived from $f_{\rm{gal}}(\textbf{v})$, the velocity distribution in the Galaxy frame according to the relations:
\begin{align}
    f_{\rm{lab}}(\textbf{v}) = f_{\rm{gal}}(\textbf{v}+\textbf{v}_{\bigodot}+\textbf{v}_{\bigoplus})
    \label{Eq:v_dist}
\end{align}
where $\textbf{v}_{\bigodot}=(11.1, 247.24, 7.25),\rm{km\,s^{-1}}$~\cite{10.1093/mnras/stw2759,10.1111/j.1365-2966.2010.16253.x} represents the velocity of the Sun in the Galactic rest frame, and $\textbf{v}_{\bigoplus}$ is the velocity of Earth as it orbits the Sun. Neglecting the modulation effect (i.e., the time dependence of Earth's velocity), we set $\textbf{v}_{\bigoplus}=(29.4,-0.11,5.90),\rm{km\,s^{-1}}$, with a magnitude ${v}_{\bigoplus}=29.79\,\rm{km\,s^{-1}}$.


Since the Gaia data reveals that the stellar halo in the neighborhood of the solar system is composed of different substructures, it is reasonable to assume that there exist DM counterparts corresponding to these substructures. Therefore, the DM halo around the solar system is not simply a pure isotropic Maxwellian distribution. Even though there is no direct evidence for the velocity distributions of the DM counterparts, it is reasonable to assume that the DM halo shares the same velocity distribution as the stellar halo. In such a situation, we must take into account other components. Consequently, the final velocity distribution of DM in the presence of substructures can be expressed as,
\begin{align}
    f_{\rm{gal}}(\mathbf{v}) = \sum_i\eta_\xi f^{\xi}_{\rm{gal}}(\mathbf{v}) 
\label{Eq:distr_gal}
\end{align}
where $\eta_{\xi}$ is the fractional contribution of the substructure $\xi$ to the total DM halo and satisfies the normalization condition $\sum_\xi \eta_\xi=1$.The function $f^{\xi}{\rm{gal}}(\boldsymbol{v})$ represents the DM velocity distribution in the substructure $\xi$, and its general form is given by:
\begin{align}
\label{Eq:distribution}
  f^{\xi}_{\rm{gal}}(\mathbf{v}) = \frac{1}{(8 \pi^3 \det{\boldsymbol{\sigma}_\xi^{2}})^{1/2} N_{_\xi,\rm{esc}}}  \exp\left(-(\mathbf{v} - \mathbf{v}_\xi)^T\frac{1}{2(\boldsymbol{\sigma}_\xi)^{2}}(\mathbf{v} - \mathbf{v}_\xi)  \right)\Theta (v_{\rm{esc}} - |\mathbf{v}|).
\end{align}
Here, $N_{\xi,\rm{esc}}$ is the normalization factor ensuring that the distribution integrates to 1, while $\mathbf{v}\xi$ and $\boldsymbol{\sigma}_\xi$ denote the mean velocity of the substructure $\xi$ and the velocity dispersion matrix, respectively, $v_{\rm{esc}}=528\,\rm{km\,s^{-1}}$ is the galactic escape speed~\cite{10.1093/mnras/stz623}.

As described in ~\cite{Buch:2020xyt}, the dominant DM substructure is the isotropic SHM, followed by the Gaia Sausage, with the mean velocity of these two substructures being zeros, and these two substructures are expected to take the dominant account of the total DM distribution with $\eta_{\rm{SHM}} + \eta_{\rm{Sausage}}\simeq 80\%$. The velocity distribution of SHM could be described in the spherical coordinate, which is
\begin{align}
    f_{\rm{SHM}} = \frac{1}{(2\pi\sigma_v^2)^{3/2}N_{\rm{SHM}}}\exp{\left(-\frac{|\mathbf{v}|^2}{2\pi\sigma_v^2}\right)}\Theta (v_{\rm{esc}} - |\mathbf{v}|).
\end{align}
with the velocity dispersion $\sigma_v=166.17\,\rm{km\,s^{-1}}$~\cite{Evans:2018bqy}.

The next important component is the Gaia Sausage DM halo, which is radially anis-tropical, the DM velocity distribution is also described by spherical coordinates,
\begin{align}
     f_{\rm{GS}} = \frac{1}{(2\pi)^{3/2}\sigma_r \sigma_{\theta} \sigma_{\phi} N_{\rm{GS}}}\exp{\left(-\frac{|\mathbf{v_r}|^2}{2\sigma_r^2}-\frac{|\mathbf{v_{\theta}}|^2}{2\sigma_{\theta}^2}-\frac{|\mathbf{v_{\phi}}|^2}{2\sigma_{\phi}^2}\right)}\Theta (v_{\rm{esc}} - |\mathbf{v}|)
\end{align}
with the velocity dispersions $\sigma_r = 262.73\,\rm{km\,s^{-1}}$ and $\sigma_{\theta} = \sigma_{\phi} = 83.09\,\rm{km\,s^{-1}}$~\cite{Evans:1996ia}.


In this work, we consider other 6 DM substructures except SHM and Gaia Sausage, the parameters of the velocity distributions are given in Table~\ref{tab:distribution}. 


\begin{table}[]
    \centering
    \begin{tabular}{|c|c|c|}
    \hline
        name & $(v_r, v_{\phi}, v_{z})$ [km/s] & $(\sigma_r, \sigma_\phi, \sigma_z)$ [km/s]\\
        \hline
        S1  & $(-34.2,-306.3,-64.4)$   & $(81.9,46.3,62.9)$ \\
        S2a & $(5.8,163.6,-250.4)$     & $(45.9,13.8,26.8)$ \\
        S2b & $(-50.6,138.5,183.1)$    & $(90.8,25.0,43.8)$ \\
        \hline
        Rg5a &  $(6.4,-74.5,-159.5)$     & $(32.4,17.5,31.7)$ \\
        Rg6b &  $(-233.2,-221.8,51.6)$   & $(32.7,14.4,115.7)$ \\
        \hline
        Cand10 & $(-37.4,20.0,192.3)$     & $(161.5,18.2,195.0)$ \\
        Cand13 &  $(-2.1,-13.2,202.2)$     & $(215.7,28.1,215.9)$ \\
        \hline
    \end{tabular}
    \caption{The velocity distribution parameters for the DM substructure we considered in this work. Except for the SHM and Gaia Sausage, we consider the two most subdominant DM substructures, two retrograde and two prograde dark shards, with the parameters selected from ~\cite{OHare:2019qxc}.}
    \label{tab:distribution}
\end{table}

In the calculation of event rate Eq.~\ref{Eq:rate}, we should get the DM halo velocity distribution in the Galaxy frame according to Eq.~\ref{Eq:distr_gal}, and then transform it into the laboratory frame according to Eq.~\ref{Eq:v_dist}. 

\begin{figure*}[htbp]
\centering
\includegraphics[width=.45\textwidth]{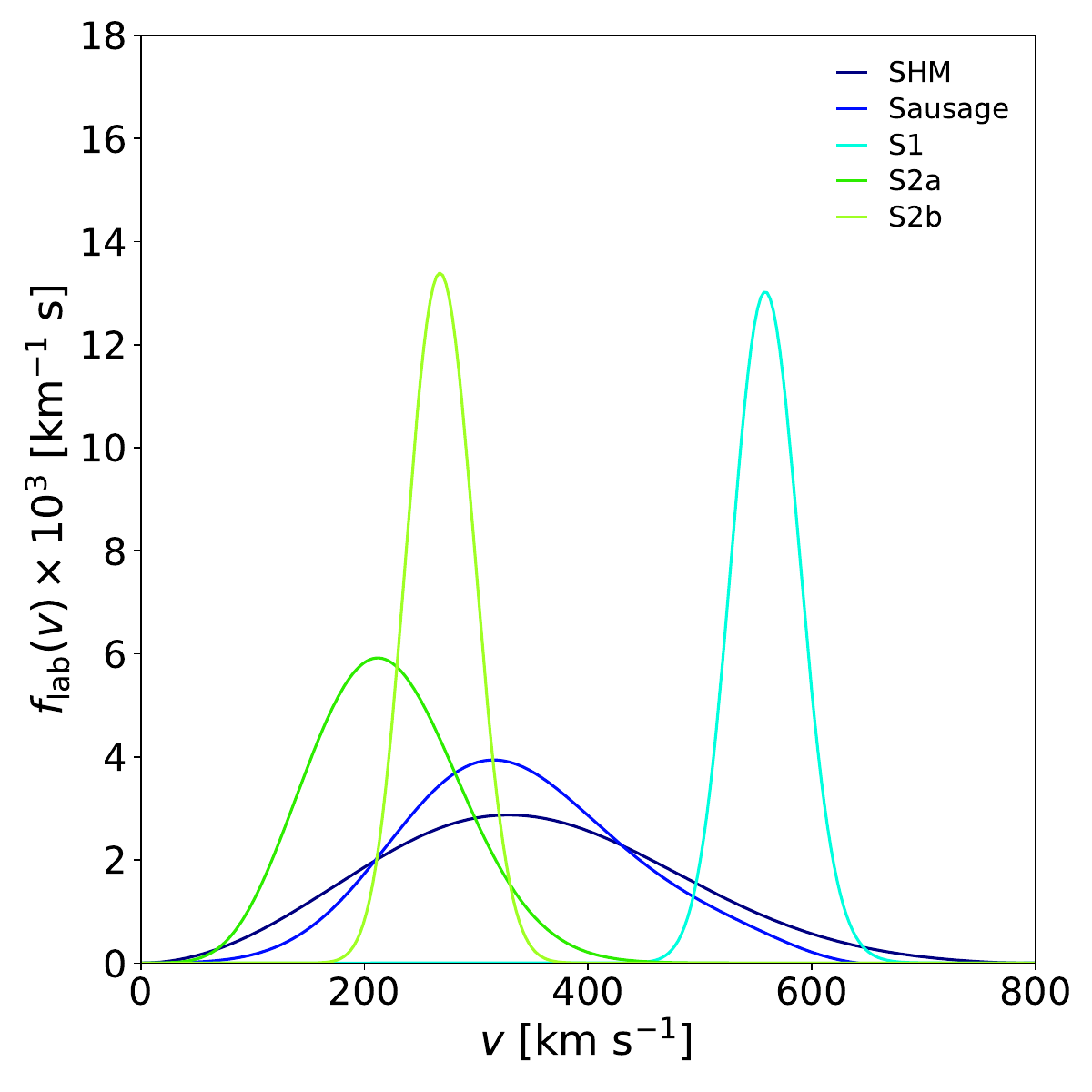}
\includegraphics[width=.45\textwidth]{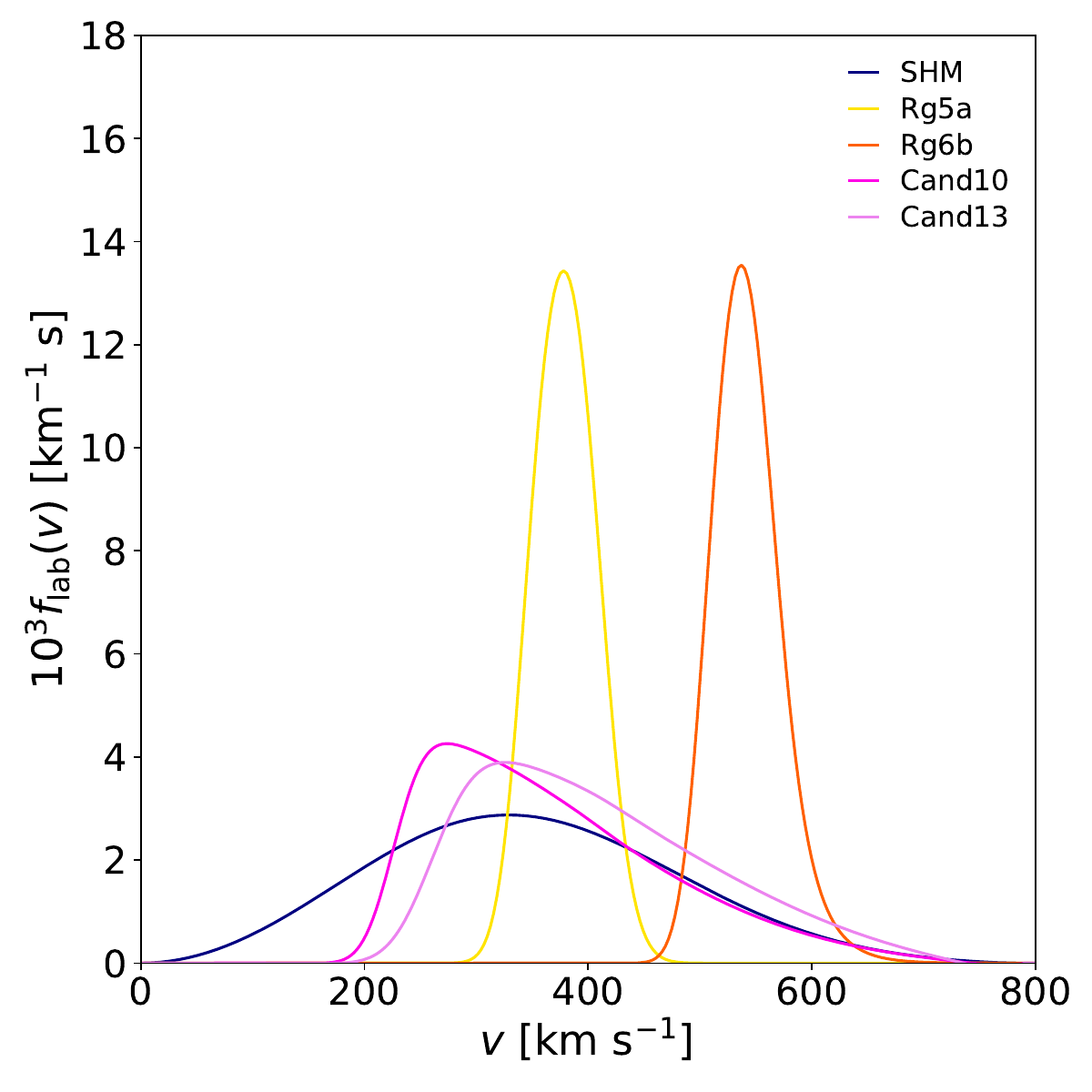}
\caption{The speed distributions for the DM substructures we considered. Compared with the SHM, some of the DM substructures such as S1, Rg6b have a large $v_{mp}$ and are more centralized around $v_{mp}$, which will have some important impact on the phonon DM detection.\label{fig:fv}}
\end{figure*}

We show the speed distributions $F(v) = v^2\int d\Omega f(\mathbf{v})$ for different DM halos and dark shards in Fig.~\ref{fig:fv}. From Fig.~\ref{fig:fv}, we can see that the speed distribution of S1 stream has a large most probable speed $v_{mp}\simeq 600\,\rm{km s^{-1}}$, this is because it is a retrograde stream, so the DM of the substructure has a large speed for the motion of the solar system in the galaxy rest frame. For the S2a and S2b streams, they are the prograde ones, so they have a small speed in the laboratory frame. Both S1 and S2 have a small dispersion because the variance of the velocity distribution in each direction is very small (less than 100 $\rm{km s^{-1}}$) compared with that of SHM and Gaia Sausage.

From the speed distribution $F(v)$, the event rate is rewritten as,
\begin{align}
        R =  \frac{1}{4\pi \rho_T} \frac{\rho_\chi}{m_\chi}  \frac{\sigma_p}{\mu_\chi^2}  \int \! dv\, F_{\rm{lab}}(v)v \int\displaylimits_{q_-}^{q_+} \! dq \,  \int\displaylimits_{\omega_\mathrm{th}}^{\omega_+} \! d\omega \,   q \, F^2_{\rm{DM}}(q) S(q, \omega),
    \label{Eq:rate_speed}
\end{align}

The speed distribution will leave an important impact on the event rate, for the integral range of the transfer momentum $q$ and the transfer energy $\omega$  directly related to the speed, from Eq.~\ref{eq:int_omg} and Eq.~\ref{eq:int_q}. 

As is shown in Fig.~\ref{fig:S_plot}, the integral region of two DM substructure cases, the S1 stream and S2a stream, with the solid line representing the region for the most probable speed $v_{mp}$, the dotted and dashed lines represent the upper $v_{up}$ and lower $v_{low}$ bound of 1-$\sigma$ region around $v_{mp}$ for certain speed distribution of DM substructure. It is obvious that S1 stream has a larger integration region for $v_{mp}$, $v_{low}$ and $v_{up}$ compared with that of S2a stream for both harmonic and anharmonic cases, which means the DM substructure speed distribution with larger $v_{mp}$ is expected to generate more events. On the other hand, the energy threshold will move the lower bound horizon line in the Fig.~\ref{fig:S_plot}, if the threshold is high enough, for example, $\omega_{th}\simeq 500\,\rm{meV}$, DM substructures with low $v_{mp}$ and small speed dispersion are difficult to leave phonon signal in the lattice.

\section{Numerical Results and Discussions}

By substituting the structure factor Eq.~\ref{eq:sfn} into the event rate Eq.~\ref{Eq:rate_speed}, it can be demonstrated that the event rate is a summation of final states with different phonon numbers. To examine the relationship between the event number and anharmonic effect from the DM substructure, we have set the DM form factor to $F(q)=1$ and the DM-nucleus scattering cross section to be equal to $1\times 10^{-42}$ cm$^2$. 

From the result in Fig.~\ref{fig:step} we show the yearly scattering event per kg-year for pure DM substructure in Table.~\ref{tab:distribution}. The results are presented for two distinct DM mass scenarios: a light case with $m_{\chi}=20\,\rm{MeV}$ and a heavy case with $m_{\chi}=80\,\rm{MeV}$. From the light case, it is evident that the DM substructure S1 and Rg6b will generate more events in the final state with any number of phonons. This is because both distributions have a larger $v_{mp}$ and their speed distributions are considerably more centralized. The substructure S1 and Rg6b generate a greater number of events at phonon number $n\simeq15$. This is due to the fact that for a large $v_{mp}\simeq 570\,\rm{km s^{-1}}$, the expected number of phonon number is $\Bar{n}\simeq m_{\chi}\simeq 13$. It indicates that the event rate is maximized at this value of the phonon number. In the large DM mass case, before the expected maximum phonon number is reached, the anharmonic case produces a lower number of events than the harmonic case, this is because the anharmonic effect will always inclined to generate events with more phonons, resulting in fewer events in low phonon number regions.

\subsection{Data Analysis Method}

\begin{figure*}[htbp]
\centering
\includegraphics[width=.45\textwidth]{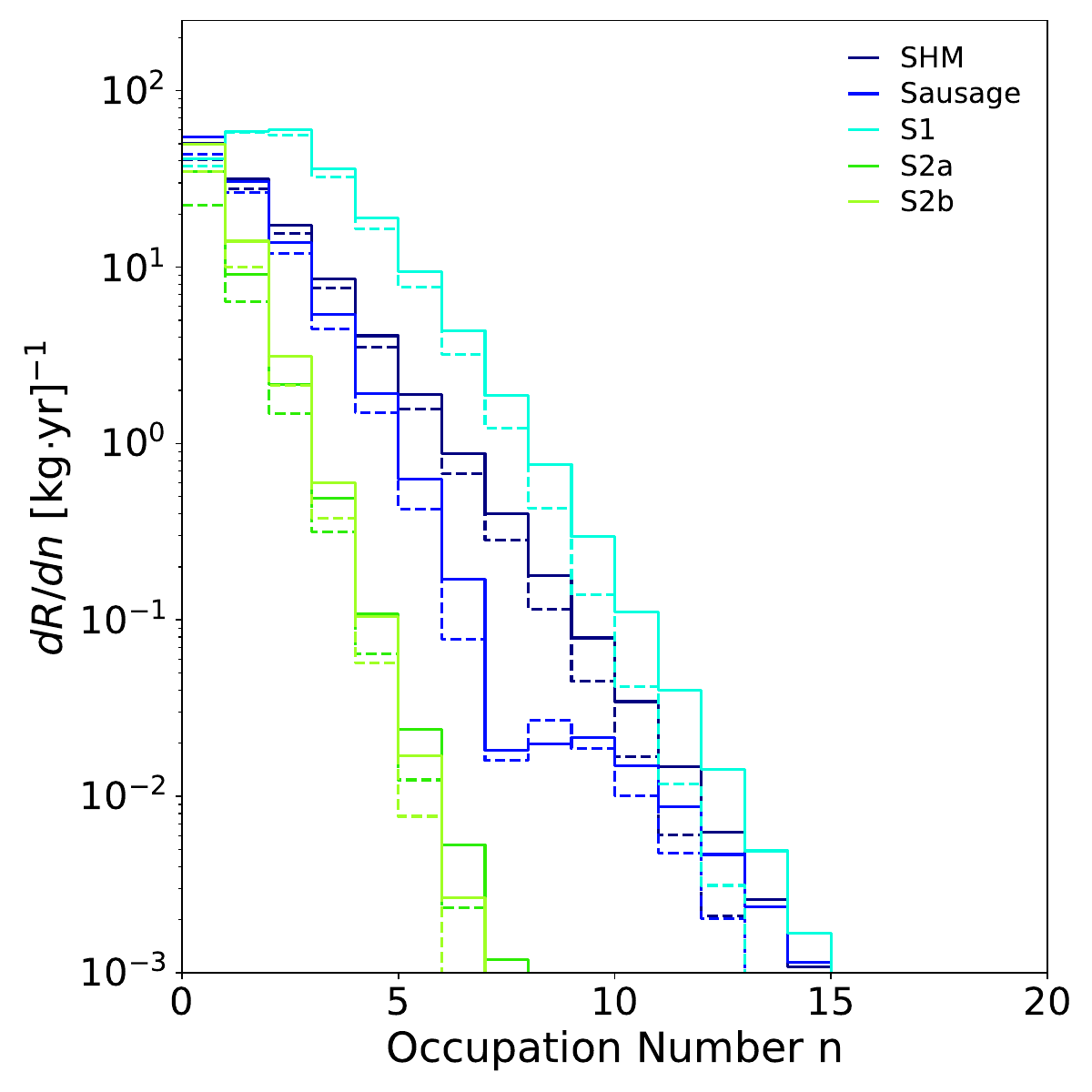}
\includegraphics[width=.45\textwidth]{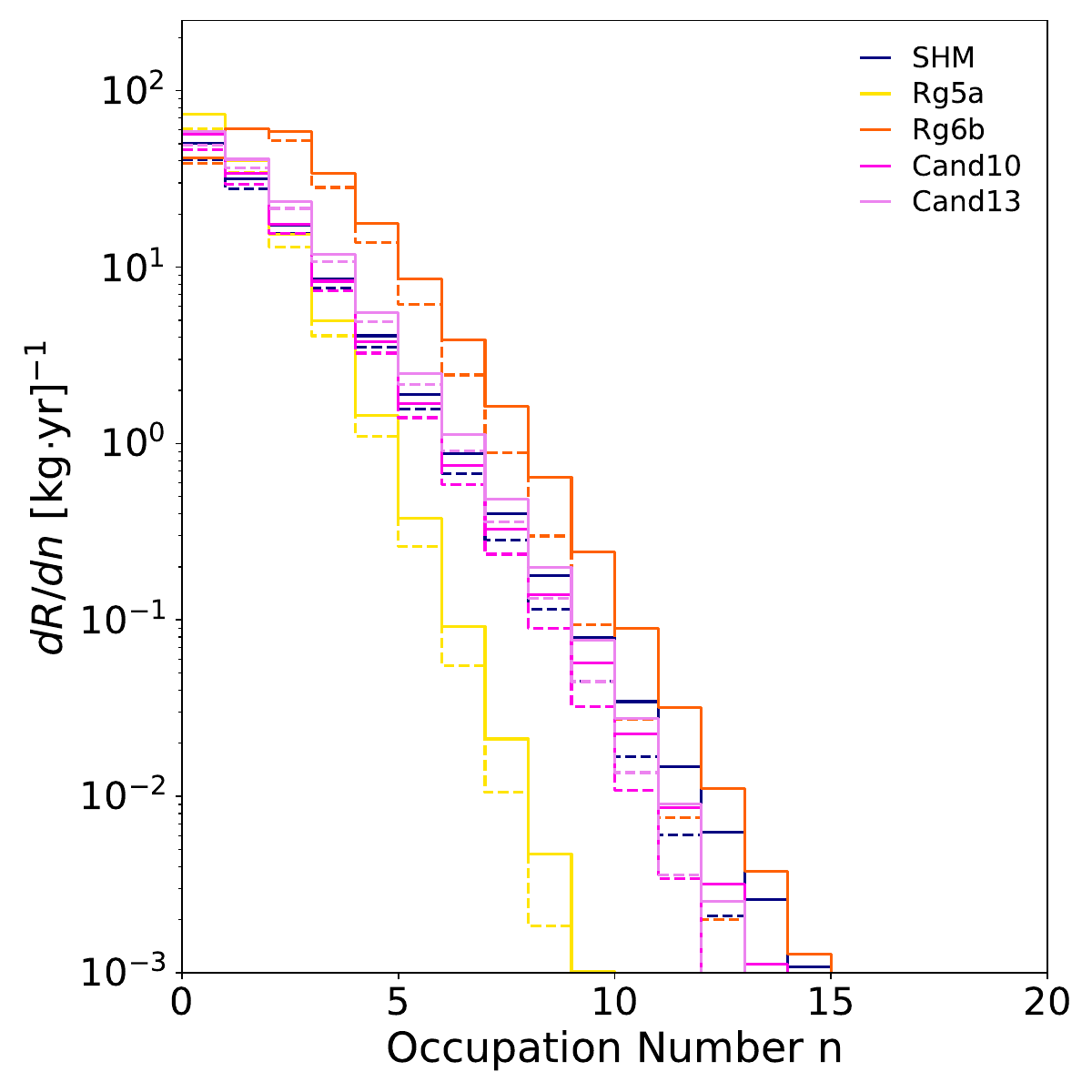} \\
\includegraphics[width=.45\textwidth]{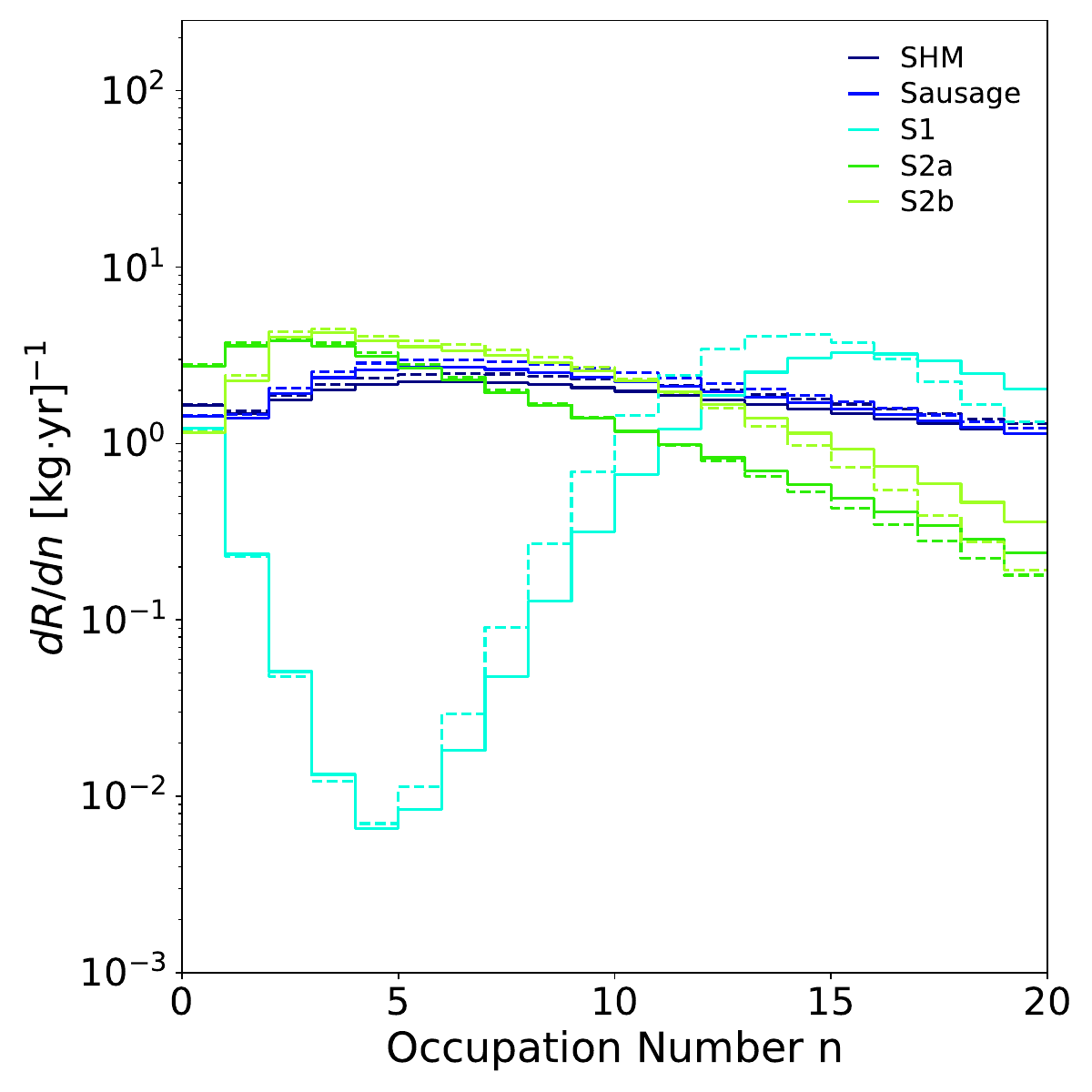}
\includegraphics[width=.45\textwidth]{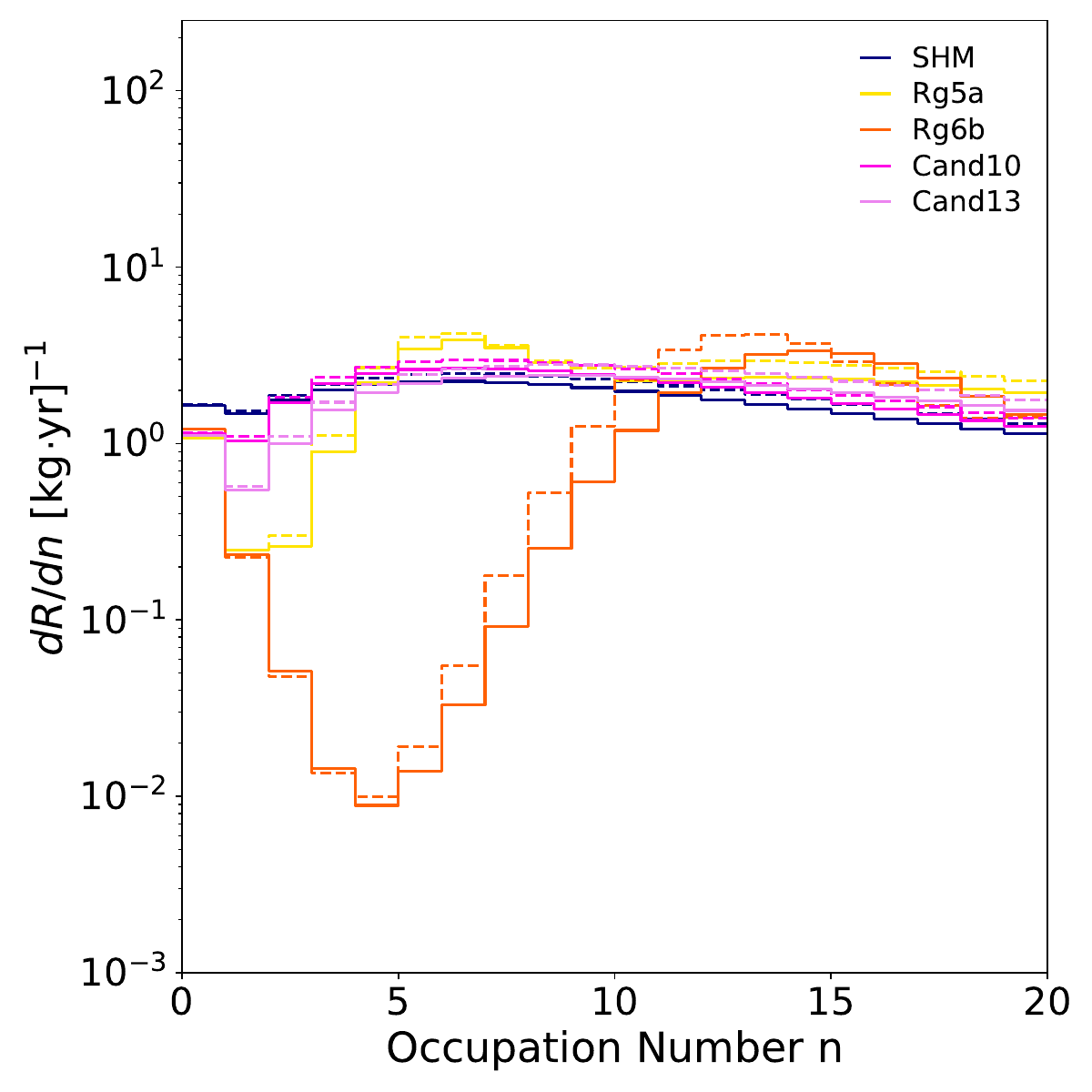}
\caption{The Quantized rate spectrum for individual n-phonon final state event, with the solid for including the anharmonic effect, and the dashed for the harmonic potential. Upper: the $m_{\chi}=20\,\rm{MeV}$ case. Lower: the $m_{\chi}=80\,\rm{MeV}$ case. In the larger DM mass case, the DM substructures S1 and Rg6b are expected to generate more events with multi-phonons, as analyzed in the main paragraph.\label{fig:step}}
\end{figure*}

The event rate has the potential to influence the sensitivity of the future phonon DM detection experiment. To study the future sensitivity of DM-nucleus scattering, it is necessary to set an exclusion line on the $m_{\chi}-\sigma_{p}$ plane. In the work of ~\cite{Trickle:2020oki,Kahn:2020fef}, it is assumed that 3 events per kg-yr were observed, and the expected scattering observation $\sigma_p$ could be observed for different DM mass $m_\chi$. However, this method is unable to be correlated with the likelihood. 

In this study, we assume that the background of the event rate per kg-yr is 3. To derive the expected sensitivity of the phonon detection experiment, we use the Asimov-Dataset~\cite{Cowan:2010js}, and we split the events into one bin only, as there are currently no datasets for the phonon detection experiment. The likelihood function is thus,

\begin{align}
    \mathcal{L}(\mathcal{D}_{Asm}|\mathbf{\theta}) = \mathcal{P}(\mathcal{D}_{Asm}|\mu_s+\mu_b)
\end{align}
where $\mathcal{P}$ is the Poisson distribution with rate $\mu_s+\mu_b$, the $\mu_s$ and $\mu_b$ are the expected event rate for the DM signal and background, with  $\mu_s(m_\chi, \sigma_p)$ is the event number per kg-yr and $\mu_{b}=3$ per kg-yr. $\mathcal{D}_{Asm}$ is the Asimov Dataset, is defined as the sum of the expected event rate for the DM signal and background, i.e., $D_{Asm} = \mu_s + \mu_b$.  


To probe the sensitivity of the phonon detection experiment, the test statistic (TS) is employed, which is defined as follows:
\begin{align}
    \mathrm{TS} = -2\ln{\left(\frac{\mathcal{L}(\mathcal{D}_{Asm}|\mu_{b})}{\mathcal{L}(\mathcal{D}_{Asm}|\mu_{s}+\mu_{b})}\right)} \thicksim \chi^2_1
\end{align}
which obeys the $\chi^2$ distribution of degree 1. To ascertain the $95\%$  confidence level (C.L.) sensitivity limit on the $m_\chi-\sigma_p$ plane,  TS is set to be $-2.71$. 

\begin{figure*}[htbp]
\centering
\includegraphics[width=.45\textwidth]{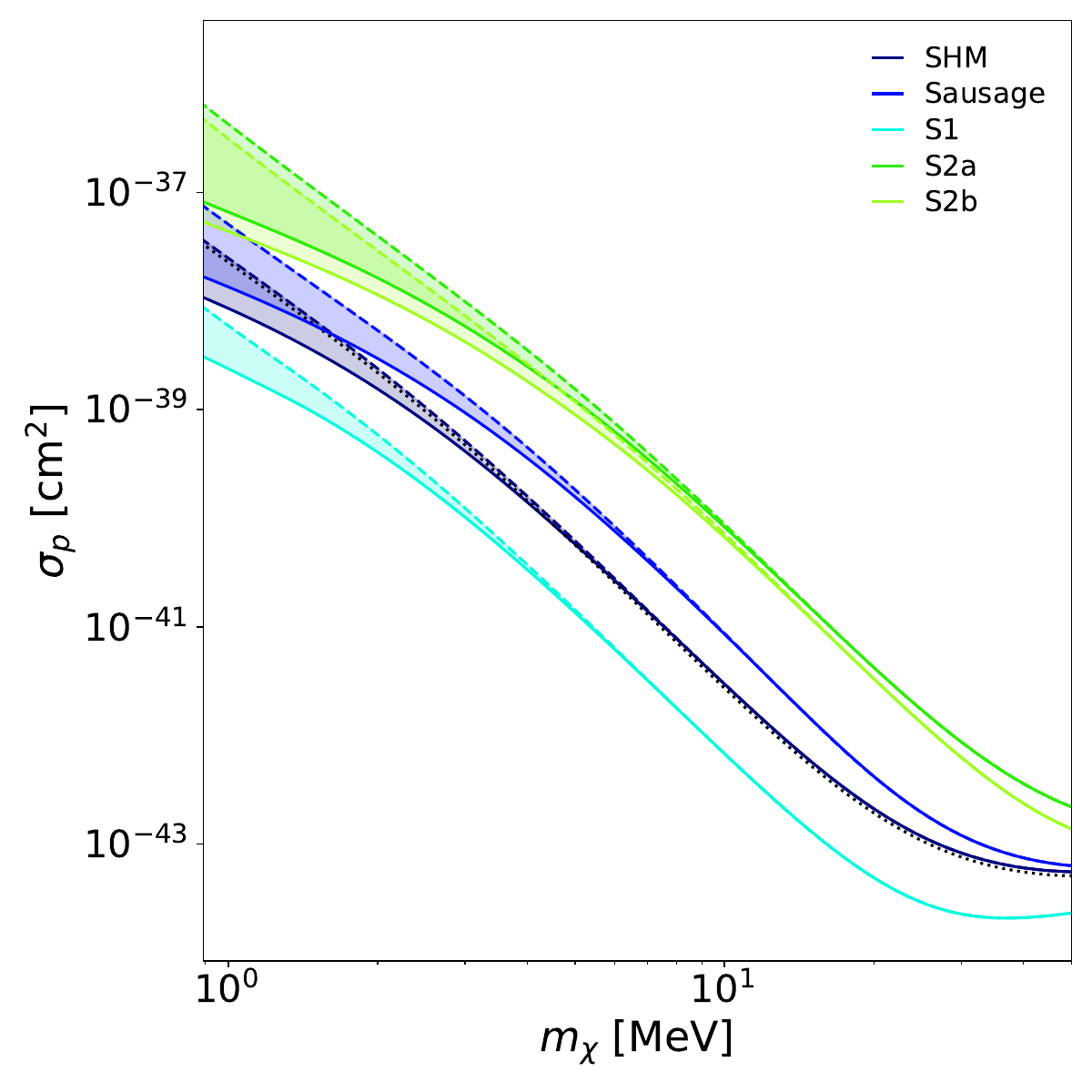}
\includegraphics[width=.45\textwidth]{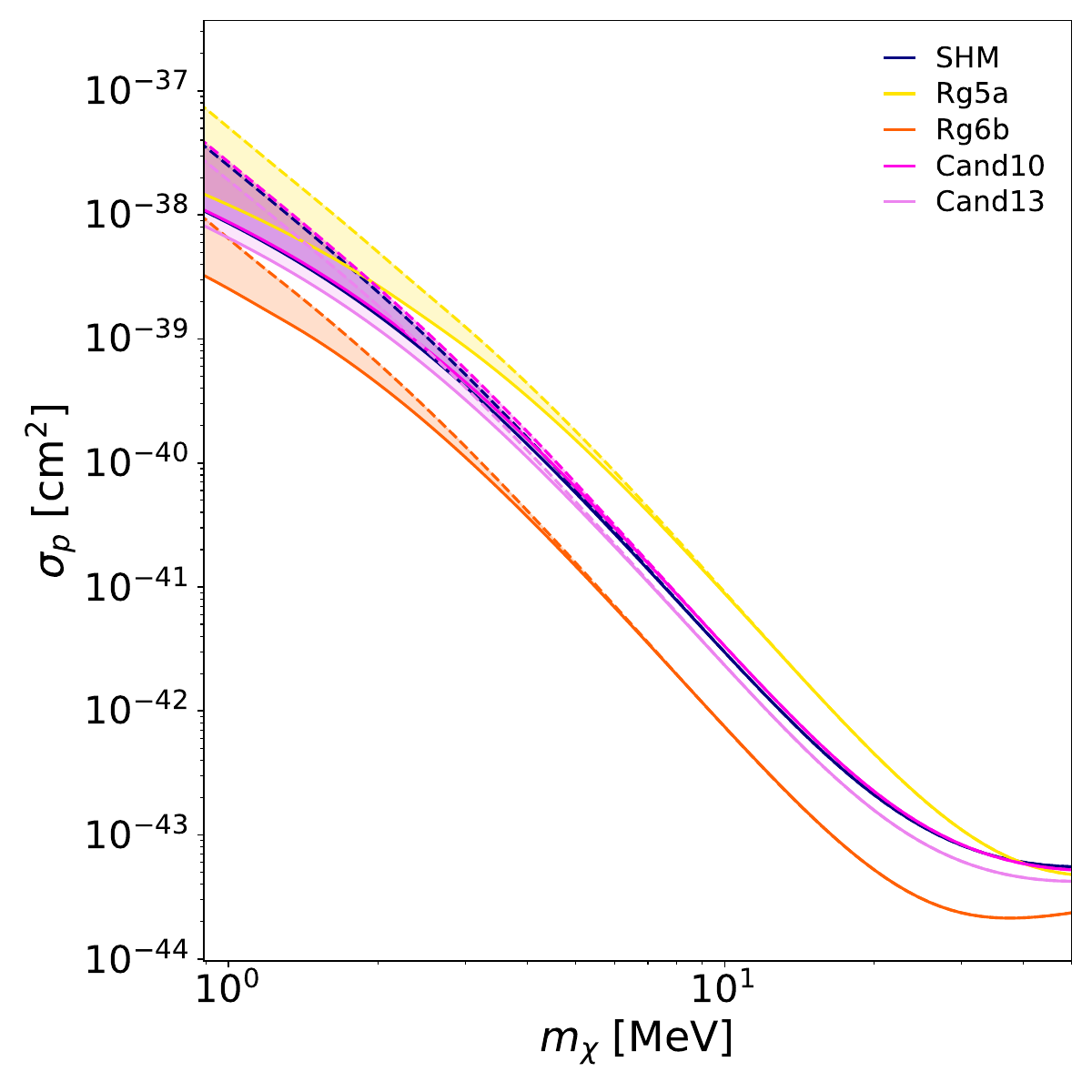}
\caption{The excluding lines for the DM substructures we considered in this work, with the solid for including the anharmonic effect, and the dashed for the harmonic potential. The 3-events limitation method for the SHM DM substructures (black dotted line) used in other works makes little difference compared with that of our paper. As analyzed in the main paragraph,  the DM stream S1 and the dark shard Rg6b set the most stringent bounds. \label{fig:excluds_pure}}
\end{figure*}

In the Fig.~\ref{fig:excluds_pure}, we show the sensitivity reach for the pure substructure cases in Table~\ref{tab:distribution}, and the energy threshold $\omega_{\rm{th}}=120$ meV. The black dotted line, derived using the method described in ~\cite{Trickle:2020oki,Kahn:2020fef}, represents the sensitivity reach in the case of pure SHM substructure. It accounts for the harmonic effect only and is identical to the result derived by our method, suggesting that our approach is solid. The remaining sensitivity reach lines demonstrate that incorporating anharmonic effects consistently yields more rigorous experimental reach estimates. Notably, DM substructures S1 and Rg6b exhibit enhanced sensitivity compared to the SHM scenario. This is directly attributable to the anticipated event rate per kilogram per year, as illustrated in Fig.~\ref{fig:step}. Furthermore, it can be observed that as the mass of the DM particle increases, the anharmonic effect becomes insignificant, except in the low-mass region, like $m_\chi\leq 3$ MeV.

\begin{figure*}[htbp]
\centering
\includegraphics[width=.45\textwidth]{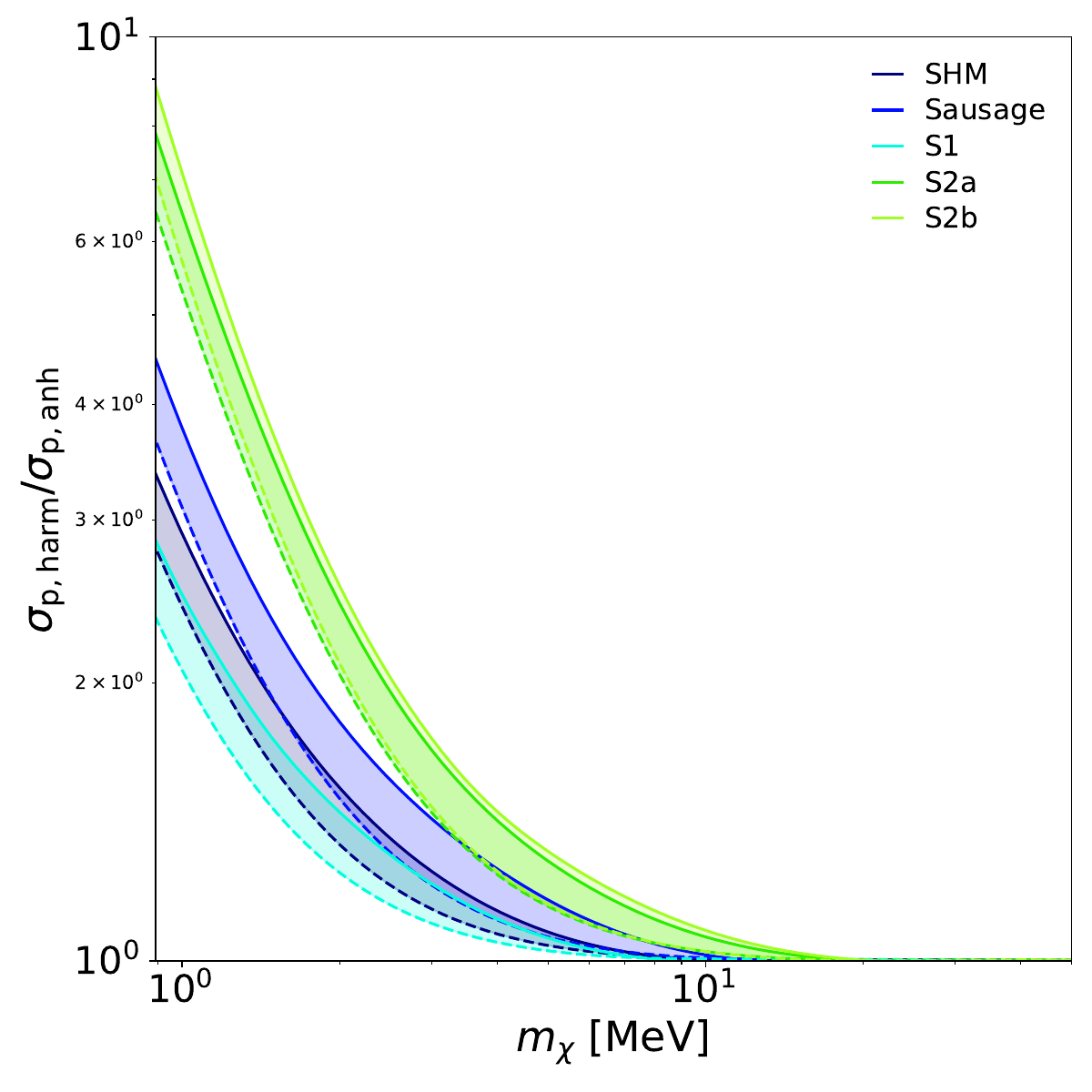}
\includegraphics[width=.45\textwidth]{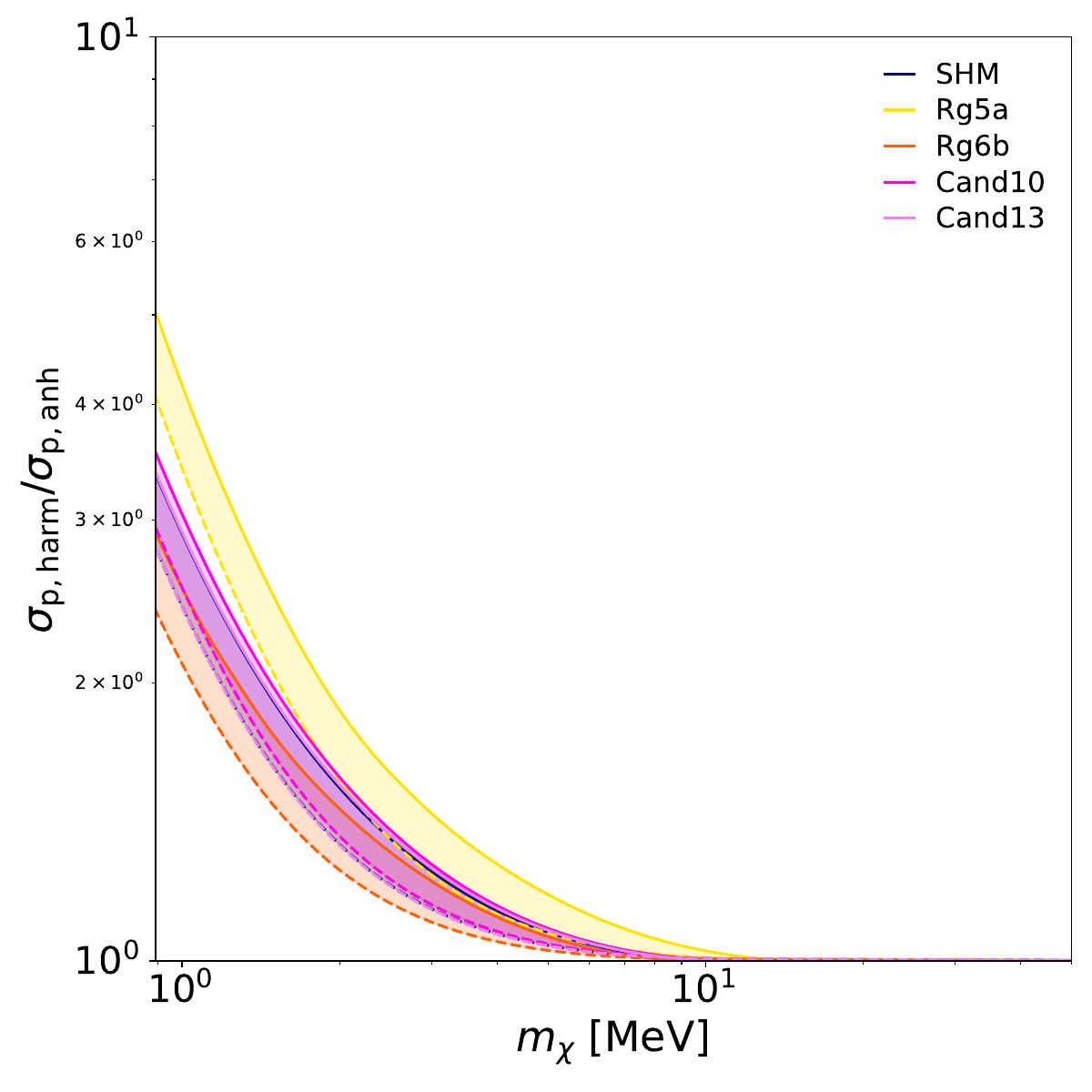}
\caption{The ratio of cross section limitations between harmonic and anharmonic. The dashed lines are for $\omega_{th}=80\,\rm{MeV}$ cases, the solid lines are for the $\omega_{th}=120\,\rm{MeV}$ cases. The limitation of considering the anharmonic effect will become more stringent. The anharmonic effect is especially pronounced in the 2a, S2b, and Rg5a DM substructure cases. \label{fig:ratio}}
\end{figure*}

In Fig.~\ref{fig:ratio}, we illustrate the ratio between the exclusion limit of anharmonic effects and that of harmonic effects. It depicts the ratio of two energy thresholds: 120 meV (solid) and 80 meV (dashed).  It is evident that an elevated energy threshold results in a more pronounced anharmonic effect. As the DM mass approaches 10 MeV, the anharmonic effect diminishes in significance. These findings align with the conclusions presented in ~\cite{Lin:2023slv}. The significance of the anharmonic effect varies depending on the specific DM substructure, as illustrated in  Fig~\ref{fig:S_plot}. In the case of S2a, S2b, and Rg5a, the anharmonic effect is particularly significant. All of these occurrences are related to the anharmonic correction to the structure factor, $S(q, \omega)$. Given that we consider the Morse potential in this work, we can simplify it as a cubic perturbation with $\lambda_3=\lambda_M$, following the analysis in ~\cite{Lin:2023slv}, the anharmonic correction to the n-phonon final state is related to the momentum transfer, when 
\begin{align}
q\lesssim \sqrt{2m_d\omega_0}\lambda_M^{k((n))}
\end{align}
the anharmonic effect can be comparable to the harmonic structure factor, where $k(n)=1$ for $n=1$, $k(n)=1/2$ for $n=2$ and $k(n)=1/3$ for other phonon numbers. Given that the momentum transfer $q\simeq 2 m_\chi v$, where $v$ is the DM velocity, as illustrated in Fig.~\ref{fig:fv}, it is evident that the most probable velocity of Rg5a is smaller than other cases, and the velocity distribution of it has a small dispersion, so the anharmonic effect is the most significant one. In future phonon experiments aimed at detecting MeV-scale DM, it can also determine which substructure is available. 

\subsection{Results}
We will apply the statistical method and phonon effects to the study of sensitivity reach for future phonon detection experiments. The parameters are the DM mass $m_\chi$, DM-proton scattering cross section $\sigma_p$, and the DM substructure fraction $\eta_{\xi}$. It is anticipated that the forthcoming experiment will be capable of detecting events at a rate of one year and one kilogram.

It is proposed that the real DM structure should contain all kinds of substructures. To achieve the desired future experiment sensitivity, it is assumed that the dominant DM halo obeys the SHM++ model~\cite{Evans:2018bqy}, whereby the dominant component is the SHM one, followed by the Gaia Sausage substructure, and then the remaining parts are other dark matter substructures as illustrated in Table~\ref{tab:distribution}. For the sake of simplicity, we assume that the remaining DM substructure comprises only a single kind. Consequently, the parameters characterizing the DM halo are given by the values of $\eta_{\rm SHM}$, $\eta_{\rm GS}$ and $\eta_{\rm DS}$. 

Given that the gravitational potential is approximately spherical ~\cite{Wegg:2018voc}, it can be inferred that the SHM and Gaia Sausage components are the dominant ones. The bound of the equipotentials ellipticity of the Milky Way ~\cite{Evans:2018bqy} and Auriga simulations of the Sausage's formation suggest that the Gaia Sausage component fraction $\eta_{\rm GS}$ could reach $20\%$. Given that the dark shard is not the dominant component of the DM halo, two cases have been set for it: $\eta_{\rm DS}=10\%$ and $\eta_{\rm DS}=20\%$. Consequently, the remaining component is the SHM one, with $\eta_{\rm SHM} = 80\% - \eta_{\rm DS}$.

It can be reasonably deduced that the DM model will also have an impact on the final discovery sensitivity. The dark photon-mediated MeV-scaled DM model plays an important role in the phenomenology, as demonstrated in~\cite{Holdom:1985ag,Fabbrichesi:2020wbt,Rizzo:2018ntg,Pospelov:2008zw,Nelson:2011sf}. This is because the dark photon is capable of coupling to both DM and SM particles through kinetic mixing, as well as an additional $U'(1)$ gauge interaction. Consequently, it is essential to consider the atomic form factor, as outlined in~\cite{Schiff:1951zza,Tsai:1973py,Emken:2019tni}, 
\begin{align}
    \left|F_{A}(q)\right|^2 = \frac{\lambda_{\rm TF}^4q^4}{(1+\lambda_{\rm TF}^2q^2)^2}
\end{align}
where $\lambda_{\rm TF}$ is the Thomas-Fermi screening length, for the Silicon detection target, $\lambda_{\rm TF}\simeq 0.37a_0=0.0928$ with $a_0$ the Bohr radius. For the momentum transfer $q\simeq 2m_\chi v$ with the velocity $10^{-3}$, the screening effect $\left|F_{A}(q)\right|^2 \simeq 0.6$ for $m_\chi=10$ MeV, giving a significant impact on the light DM with a mass around $\mathcal{O}(1)$ MeV. 
\begin{figure*}[htbp]
\centering
\includegraphics[width=.45\textwidth]{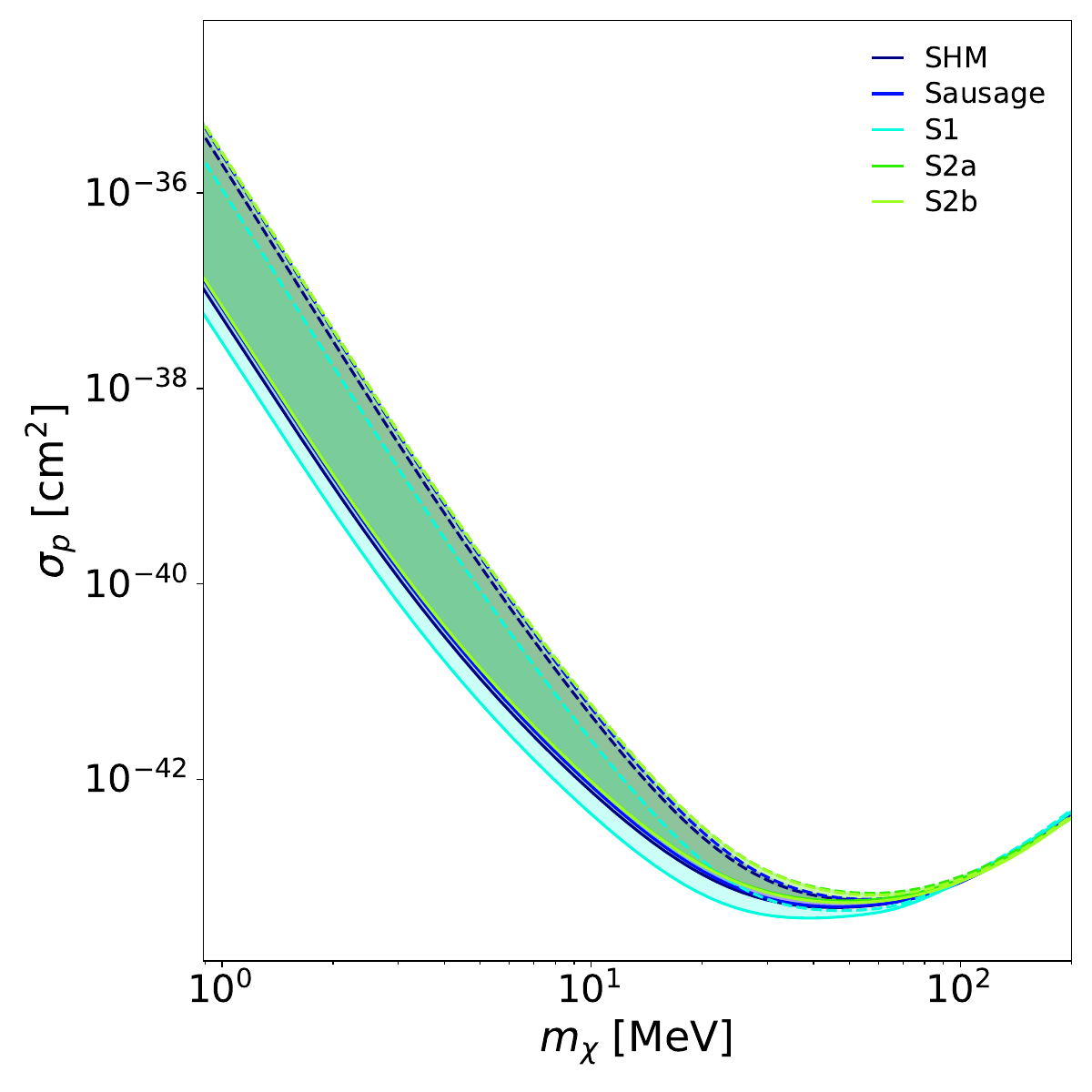}
\includegraphics[width=.45\textwidth]{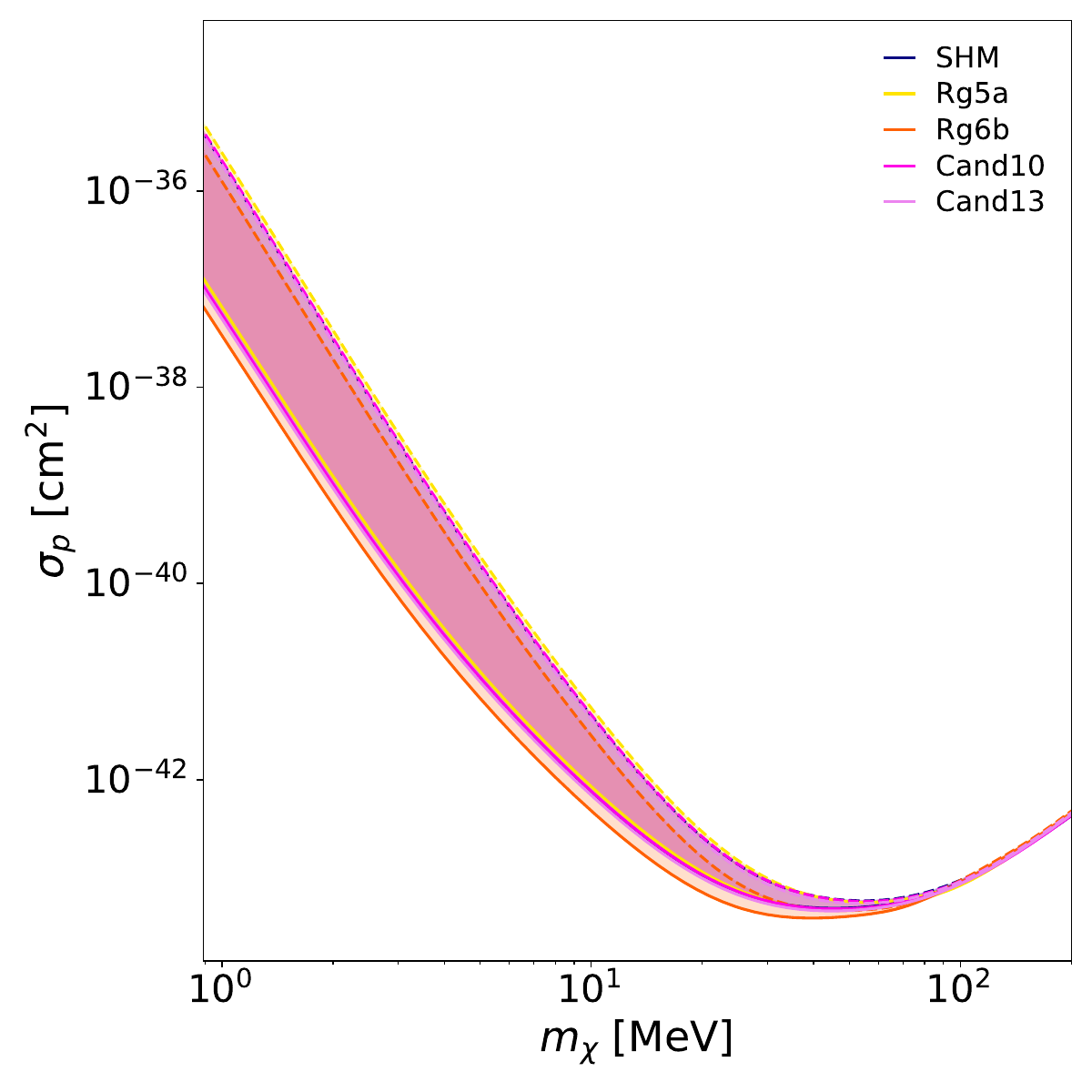}\\
\includegraphics[width=.45\textwidth]{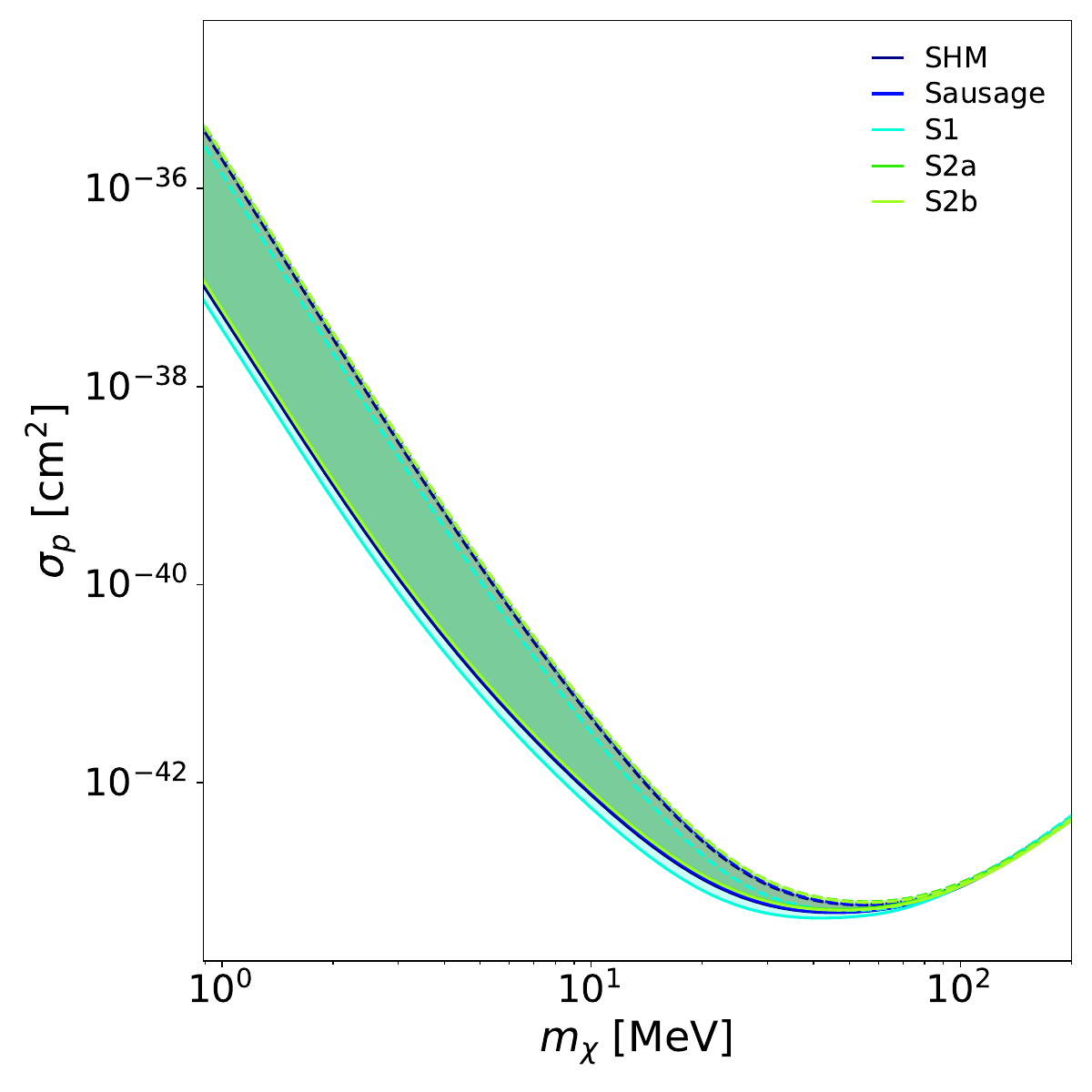}
\includegraphics[width=.45\textwidth]{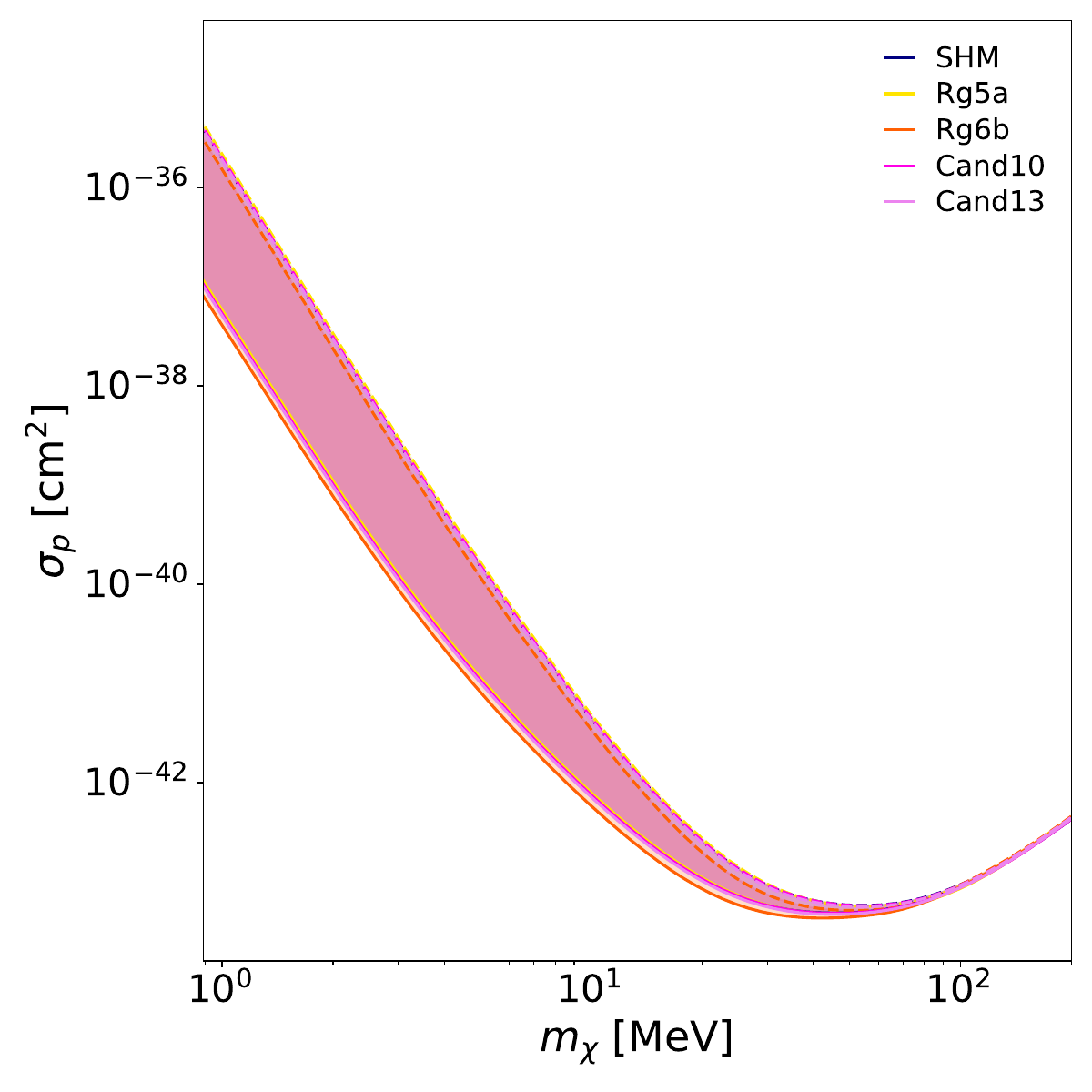}\\
\caption{The final excluding result for the DM substructure models we considered in our work, the solid lines are for $\omega_{th}=80\,\rm{meV}$, and the dashed lines are for $\omega_{th}=120\,\rm{meV}$. Upper: the modified SHM++ model plus extra substructure fraction $\eta_{\rm{DS}}=0.2$. Lower: the modified SHM++ model plus extra substructure fraction $\eta_{\rm{DS}}=0.1$. The lines labeled SHM are the SHM++ halo results.\label{fig:exclude_shmpp}}
\end{figure*}

We depict the final exclusion limits in Fig.~\ref{fig:exclude_shmpp}, with two threshold cases: $\omega_{th}=80$ meV (solid) and the $\omega_{th}=120$ meV (dashed), and the fraction of the dark shard are $\eta_{\rm{DS}}=0.2$ and $\eta_{\rm{DS}}=0.1$. The difference of limitations between the SHM++ model and other cases becomes smaller as $\xi_{\rm{DS}}$ reduces. Compared with the SHM++ halo, the dark shards S1 and Rg6b make the experimental reach more sensitive in the low DM mass region $m_{\chi}\lesssim 80\,\rm{MeV}$, and become less stringent for the heavy region $m_\chi\gtrsim 80$ MeV, for the others dark shards, the results are opposite, and these results are consistent with the analysis in the previous sections. We can also find the energy threshold becomes less important when the DM mass is $m_\chi>100$ MeV, since when the DM mass becomes larger, the deposited energy will become so large $E_d > E_r\gg \omega_0$ that events with large phonon number $\Bar{n}\simeq 2m_{\chi}v_{mp}/\omega_0$ is preferred, as the $\omega_{th}\simeq 2\omega_0$, the events number between $[0, \omega_{th}]$ is negligible compare with the rest events number.

\section{Conclusion}

In this study, we have comprehensively investigated the implications of the anharmonic nature of crystals and DM substructures on the sub-GeV DM direct detection. By treating the scattered nucleus's final state as harmonic oscillators and using the Morse potential to account for anharmonicities, we show that the anharmonic nature of the crystal is vital in determining the dark matter-nucleus scattering cross-section, especially in the low dark matter mass region, and is intricately linked to the velocity distribution of the dark matter halo, which is affected by potential substructures as indicated by Gaia satellite observations. Compared to the Standard Halo Model, certain DM substructures like S1 and Rg6b exhibit a higher most probable speed and greater centralization around it, leading to an increased event rate and prevalence of multi-phonon events, thus enhancing the experimental sensitivity. Conversely, for substructures such as S2a, S2b, and Rg5a, the lower $v_{mp}$ makes the anharmonic effect more important as the momentum and energy transfer into regions where the anharmonic potential's structure factor is significant. By incorporating these factors and utilizing likelihood analysis with Asimov datasets, we find that the expected sensitivity can be changed by a factor of 2-3 as a comparison with the prediction of SHM. Thus, we emphasize the importance of considering these effects due to the unprecedented precision in future direct detection experiments.

\acknowledgments

JG was supported by the National Natural Science Foundation of China under Grant No. 12305111, and Jiangxi Provincial Department of Education Scientific Research Program No. GJJ2200375. LW was supported by the National Natural Science Foundation of China under Grant No. 12275134. BZ was supported by the National Natural Science Foundation of China under Grant No. 12275232

\vspace{-.3cm}
\bibliography{mybibliography}

\end{document}